\documentclass[showpacs,showkeys,preprintnumbers]{revtex4}
\usepackage{graphicx}

\newcommand{\be}{\begin{equation}}
\newcommand{\ee}{\end{equation}}
\newcommand{\bea}{\begin{eqnarray}}
\newcommand{\eea}{\end{eqnarray}}
\newcommand{\Tr}{{\rm Tr}}

\begin{document}
\title{Numerical computation of the beta function
  of large $N$ SU($N$) gauge theory coupled to an adjoint Dirac fermion}
\date{\today}
\author{A. Hietanen}
\email{hietanen@cp3-origins.net}
\affiliation{ 
CP3-Origins and the Danish Institute for Advanced Study DIAS,
University of Southern Denmark, Campusvej 
 55, DK-5230 Odense M, Denmark.}
\author{R. Narayanan}
\email{rajamani.narayanan@fiu.edu}
\affiliation{ 
Department of Physics, Florida International University,
Miami, FL 33199, USA.}

\begin{abstract}
We use a single site lattice in four dimensions to study the scaling of large $N$ 
Yang-Mills field coupled
to a single massless Dirac fermion in the adjoint representation. 
We use the location of the strong to weak coupling transition 
defined through the eigenvalues of the folded Wilson loop operator to
set a scale. We do not observe perturbative scaling in the region
studied in this paper. Instead, we observe that the scale changes very slowly with the bare
coupling.
The lowest eigenvalue of the overlap Dirac operator is another scale that
shows similar behavior as a function of the lattice coupling.
We speculate that this behavior is due to the beta function appoaching close
to a zero.
\end{abstract}


\preprint{CP3-Origins-2012-009, DIAS-2012-10}

\maketitle

\section{Introduction}
Particle accelerators experiments provide strict bounds for the beyond standard model physics. For technicolor it means that the coupling constant has to exhibit walking behavior. Otherwise the theory cannot simultaneously explain the mass pattern of standard model fermions and the suppression of the flavor changing neutral currents  \cite{Holdom:1981rm,Yamawaki:1985zg,Appelquist:1986an,Andersen:2011yj}. Hence, lattice studies of vector like gauge theories with appropriate choice of fermion matter with the aim of understanding the conformal window has recently attracted considerable attention(see~\cite{DelDebbio:2010zz} and references therein). The gauge group is chosen to be SU(N) and  the number and representation of fermions is such that the theory is expected to be conformal or near conformal \cite{Sannino:2004qp}. 

Let 
\be
b=\frac{1}{g^2N}
\ee
define the inverse 't Hooft coupling on the lattice. 
Let 
\be
t=\ln a
\ee
define the logarithm of a lattice scale where $a(b)$ could be the square root of the
string tension measured on the lattice at the coupling $b$.
The beta function of the lattice is defined as
\be
\beta(b) = \frac{db(t)}{dt}.
\ee
The perturbative beta function leads off as
\be
\beta(b) = -b_0 - \frac{b_1}{b} + \cdots.
\ee
As is well known~\cite{Weinberg:1996kr}, only the one and two loop
coefficients, $b_0$ and $b_1$, are
universal and the higher order coefficients in a Taylor expansion
of $\beta(b)$ in powers of $b^{-1}$ depend on the choice of $a(b)$. In
fact, one can imagine choosing an $a(b)$ such that all higher order
coefficients are zero. We will not have such control on the choice
of $a(b)$. In particular, there is no reason to expect the location of
the zero of the
beta function to be independent of the choice of $a(b)$ -- all we can
expect
is for the zero to remain stable if it is at a perturbatively weak coupling.

The choice of fermionic matter can be motivated by the presence of a zero 
in the two-loop perturbative beta function.
In order to maintain asymptotic freedom, all choices are
such that $b_0>0$. The two loop beta function
has a zero if $b_1<0$. Some of the choices currently being investigated are:
\begin{itemize}
\item SU(3) gauge group with twelve Dirac flavors of
  fermions in the fundamental
  representation~\cite{Hasenfratz:2011xn,Appelquist:2011dp,Fodor:2011tu} --
$b_1$ is negative if we have nine or more Dirac flavors but the
zero of the two loop beta function occurs at smaller coupling for
larger flavors.
\item SU(2) gauge group with two Dirac flavors of fermions in the
  adjoint
  representation~\cite{Catterall:2011zf,DeGrand:2011qd,Hietanen:2009az,Bursa:2011ru} --
This is the only choice based upon $b_0$ and $b_1$ since
$b_0 < 0$ if we choose three or more Dirac flavors and $b_1>0$
if we choose one Dirac flavor.
\item SU(3) gauge group with two Dirac flavors in the two-index
  symmetric
  representation~\cite{Shamir:2008pb,Fodor:2011tw,Kogut:2011ty} -- In this case
  $b_1>0$ if there is only one Dirac flavor. One can also choose
  three Dirac flavors and still maintain asymptotic freedom.
\end{itemize}

The case of SU(N) gauge theory coupled to $f$ flavors of Dirac
fermions in the adjoint representation  is
interesting for two reasons:
\begin{itemize}
\item The first two coefficients of the beta function are
\be
b_0= \frac{11-4f}{24\pi^2};\ \ \ b_1=\frac{17-16f}{192\pi^4},
\ee
and are independent of $N$~\footnote{Trivial coefficients of $N$ get
  absorbed since we have used the 't Hooft coupling and not $g^2$.}.
The three interesting choices for a theory with an infra-red
fixed point are $f=\frac{3}{2},2,\frac{5}{2}$ based on the two-loop beta function.
\item Numerical evidence along with continuum
  arguments~\cite{Bringoltz:2011by}-\cite{Kovtun:2007py}
suggest that Eguchi-Kawai reduction holds in
  the large $N$ limit as long as one uses periodic boundary conditions for
fermions. This is expected to be the case for $f\ge \frac{1}{2}$~\cite{Hietanen:2009ex} and
for non-zero quark masses \cite{Hietanen:2010fx}. 
\end{itemize}

We have the possibility to study theories with an infra-red
fixed point that have only four $SU(N)$ degrees of freedom provided we
consider the $N\to\infty$ limit. For finite $N$, the massless fermionic
operator
is a finite dimensional operator that decouples into chiral sectors.
The fermion determinant is positive in each chiral sector and we can
define a theory for any real value $f$ since we can write
\be
\left(\det \not\!\!D \right)^f  = e^{f\ln\det\ \not D}.
\ee
If $\frac{11}{4} > f >\frac{17}{16}$, the two loop beta function has a zero and the
theory has an infra-red fixed point.

Our aim in this paper is to use
overlap fermions~\cite{Edwards:1998wx}-\cite{Narayanan:1994gw} and
study the $f=1$ theory on a single site lattice. 
We do not expect the beta function to have a zero from the
perturbative viewpoint. Even if it has a zero, we expect it to be at
strong
coupling. With this in mind we expect a computation of the
running
coupling to agree with the two-loop running. Contrary to this expectation, we will show that the coupling
runs
much faster than what is expected from perturbative running.

The model on the single site lattice is
defined in Sec.~\ref{model}. We will numerically study this model
using the Hybrid Monte Carlo (HMC)
algorithm with pseudofermions as described in Sec.~\ref{algorithm}.
It is numerically difficult to extract the string tension. On the
other hand there is an observable based on the Wilson loop operator~\cite{Narayanan:2006rf,Narayanan:2007dv}
that shows a transition from weak to strong coupling and we will
use the location of this transition to set our scale as discussed in
Sec.~\ref{secomega}. We will also look at the eigenvalues closest to zero of the
overlap Dirac operator. We will compare the behavior of the lowest
positive eigenvalue as a function
of the lattice coupling and compare its behavior to the scale set
using the Wilson loop operator.

Results for the behavior of the scales set using the Wilson loop
operator and the lowest positive eigenvalue of the massless Dirac
operator
are discussed in detail for the case of theory with massless fermions
in Sec.~\ref{results}.
We will show that both scales are monotonic in the coupling and that
they both
vary
very slowly with the coupling.
We will speculate on the possibility of a near-zero of the
beta function in Sec.~\ref{nonanalytic}.

\section{The model}\label{model}
The action on a single site lattice with
one flavor of adjoint Dirac overlap fermion is given by
\be
S = S_g + S_f.
\ee
The gauge action is 
\be
S_g = -12bN P;\ \ \ \ 
P=\frac{1}{12}\sum_{\mu\ne\nu=1}^4
\Tr  \left [ U_\mu U_\nu U_\mu^\dagger U_\nu^\dagger\right],
\label{gaction}
\ee
where the four gauge degrees of freedom, $U_\mu$ ($\mu=1,2,3,4$), are SU($N$) matrices.
The lattice gauge coupling constant is $b=\frac{1}{g^2N}$.
The overlap fermion action is
\be
S_f = -f\ln\det H_o(\mu),
\ee
where
the Hermitian 
massive overlap Dirac operator is defined by
\be
H_o(\mu)= \frac{1}{2}\left [ \left( 1 + \mu \right)\gamma_5 +
\left(1-\mu\right)\epsilon(H)\right],\label{hover}
\ee
with $\mu\in[0,1]$ being the bare mass.
We note that the eigenvalues of $H_o(0)$ are in the range $[-1,1]$
with exact zero eigenvalues and exact $\pm 1$ eigenvalues
corresponding
to a gauge background with non-zero topology. 
The Hermitian Wilson Dirac operator 
for adjoint fermions is
given by
\bea
H &=& \pmatrix{ 4 - m -\frac{1}{2}\sum_\mu \left( V_\mu + V_\mu^t\right)
& \frac{1}{2}\sum_\mu \sigma_\mu \left(V_\mu - V_\mu^t\right) \cr
-\frac{1}{2}\sum_\mu \sigma^\dagger_\mu \left(V_\mu - V_\mu^t\right) &
-4 + m +\frac{1}{2}\sum_\mu \left( V_\mu + V_\mu^t\right)\cr}\cr
&=& (4-m)\gamma_5
- \sum_\mu \left ( w_\mu V_\mu + w_\mu^\dagger V_\mu^t\right),
\label{wilson}
\eea
where
\be
w_\mu = \frac{1}{2}
\pmatrix { 1 & -\sigma_\mu \cr \sigma_\mu^\dagger & -1\cr}
\ee
and $V_\mu$ are the link matrices in adjoint representation.
The action of $V_\mu$ on $\Phi$ is given by
\be
V_\mu \Phi = U_\mu \Phi U_\mu^\dagger;\ \ \ 
V^t_\mu \Phi = U^\dagger_\mu \Phi U_\mu.\label{vaction}
\ee

One can verify that $H$ is Hermitian in the usual sense:
\be
\Tr \Psi^\dagger H \Phi = \left[ \Tr \Phi^\dagger H \Psi \right]^* =
\Tr \left[ (H\Psi)^\dagger \Phi\right].
\ee
Therefore $\Psi^\dagger H = (H\Psi)^\dagger$ and
in addition it is also true that $\Tr H\Phi=0$ if $\Tr\Phi=0$.
The same is also true for $H_o(\mu)$.

\section{The numerical algorithm}\label{algorithm}

We used the Hybrid Monte Carlo (HMC) algorithm to generate $U_\mu$
according to the measure
\be
Z = \int [dU_\mu] e^{-S}.
\ee
Let us introduce a Hamiltonian 
\be
\mathcal{H} =  \frac{1}{2} \sum_{\mu=1}^4 \Tr H_\mu^2 + S,
\ee
where matrices, $H_\mu$; $\mu=1,2,3,4$
are elements of the $su(N)$ algebra and conjugate to $U_\mu$.
The HMC algorithm involves the computation of the force,
$\frac{\partial S}{\partial U_\mu^{ij}}$. The gauge part of the force
is simple to compute numerically, but the fermionic part of the force
is computationally intensive. An {\sl exact} algorithm was developed
in~\cite{Hietanen:2009ex}
to compute the fermionic part of the force. This algorithm scales like
$N^6$.
In addition to using this algorithm, we also developed a
pseudo-fermion
algorithm in order to compute the fermionic part of the force which
scales like $N^4$. We present the details of the pseudo-fermion
algorithm in this section. Both algorithms were used to obtain the
numerical data presented in this paper.

We note that
\be
H_{o\pm}^2(\mu) = \frac{1+\mu^2}{2}P_\pm \pm \frac{1-\mu^2}{2} P_\pm
\epsilon(H) P_\pm; \ \ \  P_\pm=\frac{1\pm\gamma_5}{2},
\label{hoversq}
\ee
and
\be
\det H_o(\mu) = \det H_{o+}^2(\mu) = \det H_{o-}^2(\mu),
\ee
in the zero topological sector.~\footnote{We are assuming that global topology
  is completely suppressed and one can restrict the theory to the zero
  topological sector.}
The overlap fermion action can be rewritten as
\be 
S_f =  \Tr \left[ \Phi^\dagger_+ \left[ H_{o+}^2(\mu)\right]^{-1} \Phi_+\right];
\label{oaction}
\ee where 
the pseudofermions $\Phi_+$ have positive chirality and 
are traceless $N\times N$ complex matrices with an additional two component spin index.
 
For numerical purposes, we will represent $\epsilon(H)$ as
\be
\epsilon(H) = \sum_{k=1}^n \frac{ r_k H}{H^2 + p_k};\ \ \  
0< p_1 < p_2 \cdots< p_n,\label{rateps}
\ee 
with $n$ chosen such that the representation is accurate 
in the spectral range of $H^2$ assuming some lower bound on 
the spectrum of $H^2$.


The algorithm starts with one choice for $U_\mu$. Then, we draw
$H_\mu$ according to a Gaussian distribution. We also draw
Dirac indexed traceless Hermitian matrices, $\Psi$ according
to the Gaussian distribution, $\Tr \Psi^\dagger \Psi$, and set
\be
\Phi_+ = P_+ H_o(\mu) \Psi.\label{phiplus}
\ee

The equations of motion for $U_\mu$ are
\be
\frac{ d U_\mu}{d\tau} = i H_\mu U_\mu.\label{ueqn}
\ee
Setting $\frac{d \mathcal{H}}{d\tau}=0$ 
results in
\be
\sum_{\mu=1}^4 \Tr \left [ H_\mu \frac{ dH_\mu}{d\tau} \right]+
\frac{ dS_g}{d\tau} + \frac{ dS_f}{d\tau}=0,
\ee
and 
\be
\frac{ dS_g}{d\tau} = \sum_{\mu=1}^4  \Tr 
\left [ H_\mu D^g_\mu\right];\ \ \ \ 
\frac{ dS_f}{d\tau} = \sum_{\mu=1}^4  \Tr 
\left [ H_\mu D^f_\mu\right].
\ee
The equation of motion for $H_\mu$ is given by
\be
\frac{dH_\mu}{d\tau} = -D_\mu^g -D_\mu^f.
\ee

Taking the derivative of $S_g$ in (\ref{gaction}) with
respect to $\tau$ and using (\ref{ueqn}) we arrive at
\be
D^g_\mu = -ibN \sum_{\nu=1}^4 \left [
U_\mu U_\nu U_\mu^\dagger U_\nu^\dagger
+ U_\mu U_\nu^\dagger U_\mu^\dagger U_\nu
- U_\nu^\dagger U_\mu U_\nu U_\mu^\dagger
- U_\nu U_\mu U_\nu^\dagger U_\mu^\dagger  \right ]
\ee

The derivative of $S_f$ in (\ref{oaction}) with respect to $\tau$
using (\ref{hoversq}) is
\be
\frac{d S_f}{d\tau} = -\frac{1-\mu^2}{2} \Tr \left [ \Upsilon^\dagger_+ \frac{
    d\epsilon(H)}{d\tau} \Upsilon_+\right];\ \ \ 
\Upsilon_+= \left[ H_{o+}^2(\mu)\right]^{-1} \Phi_+.\label{upsplus}
\ee
Substituting the representation (\ref{rateps}) for $\epsilon(H)$, we
can write
\be
\frac{d S_f}{d\tau} = -\frac{1-\mu^2}{2} \sum_{k=1}^n \left(
r_k p_k \Tr \left[ \Upsilon^\dagger_k \frac{dH}{d\tau} \Upsilon_k\right]
-r_k \Tr \left[ \Xi^\dagger_k \frac{dH}{d\tau} \Xi_k\right]\right);\label{upsdef}
\ee
where
\be
\Upsilon_k = \frac{1}{H^2+p_k} \Upsilon_+;\ \ \ \
\Xi_k = H \Upsilon_k.\label{upskdef}
\ee

Using (\ref{wilson}), (\ref{vaction}) and (\ref{ueqn}),
we can show that
\be
\Tr \left [ X^\dagger  \frac{dH}{d\tau} X\right] = \sum_{\mu=1}^4 \Tr
\left[ H_\mu A_\mu(X)\right],
\ee
where
\be
A_\mu(X)= i \sum_{i,j=1}^4 \left(
{w_\mu^\dagger}^{ij} \left[X_j, U_\mu  X_i^\dagger U_\mu^\dagger \right]
+{w_\mu}^{ij} \left[X_i^\dagger, U_\mu X_j U_\mu^\dagger\right]
\right),
\ee
for any complex matrix $X$. It is clear that
$A^\dagger_\mu(X) = A_\mu(X)$ and that $\Tr A_\mu(X) = 0$.
Therefore,
\be
D_\mu^f = -\frac{1-\mu^2}{2} \sum_{k=1}^n \left[ r_k p_k
  A_\mu(\Upsilon_k) - r_k A_\mu(\Xi_k)\right].
\ee

Given $\Phi_+$ in (\ref{phiplus}), we compute $\Upsilon_+$ in (\ref{upsplus}) with the
standard conjugate gradient algorithm. Each action of $H_{o+}^2(\mu)$
that is part of the conjugate gradient algorithm involves the action
of $\epsilon(H)$ on a Dirac indexed traceless Hermitian matrix.
We use the multiple mass conjugate
algorithm for each action of $\epsilon(H)$ represented by
(\ref{rateps}).
The core of the fermion algorithm is the action of $H$ on a 
Dirac indexed traceless Hermitian matrix and this operation scales
like $N^3$ [see (\ref{vaction})]. In addition, the computational cost
depends on the gap of $H$ and $H_{o+}(\mu)$. The former is large
and therefore does not seriously affect the computational cost.
Since we are interested in studying chiral properties of the theory
and want to work with as small a bare mass, $\mu$, as possible
the smallest eigenvalues of $H_{o+}(0)$ will scale like $N^{-2}$
and the condition number grows like $N^2$.

\section{Operators}
We will focus on measuring two quantities that will help us understand
the running of the coupling with the scale and reveal numerical
evidence for a singular point. One observable looks at the property of
the gauge field and the other looks at the property of the massless fermion.
\subsection[Weak to strong coupling transition]{Weak to strong coupling transition~\cite{Narayanan:2007dv}}\label{secomega}

A folded $L\times L$ square Wilson loop operator in the $\mu-\nu$ plane is given by
\be
W(L) = U_\mu^L U_\nu^L {U_\mu^\dagger}^L {U_\nu^\dagger}^L.\label{floop}
\ee
The eigenvalues, $e^{i\theta_k}$; $k=1,\cdots,N$ of this operator are
gauge invariant. Let $p(\theta;L,b)$ be the distribution of these
eigenvalues
with $\theta\in[-\pi,\pi)$.
This distribution undergoes a transition at $N\to\infty$ as the size,
$L$, is changed at a fixed coupling $b$: the
distribution has a gap at $\pi$ for small areas and it becomes gapless
for areas larger than a critical area $A_c(b)$.
There is a
universal
function describing the distribution in terms of the scaled variables
derived from $A(b)$ and $\theta$ in the vicinity of $A_c(b)$ and
$\pi$. 

Let
\be
O_N(\xi;L,b) = \left\langle \det \left( e^{\frac{y}{2}} + e^{-\frac{y}{2}}
    W(L)\right)\right\rangle;\ \ \ \ \xi=\tanh\frac{y}{2}.
\ee
The region close to $\xi=0$ probes $\theta$ close to $\pi$.
Let 
\be
O_N(\xi;L,b) = C_0(L,b,N) + C_1(L,b,N) \xi^2 + C_2(L,b,N) \xi^4 + \cdots.
\ee
It is useful to define a Binder cumulant type quantity
\be
\Omega(L,b,N) = \frac{ C_0(L,b,N) C_2(L,b,N)}{C_1^2(L,b,N)}.\label{omegadef}
\ee
One can show using the universal scaling function that
\be
\Omega(L_c(b),b,\infty) =
\frac{\Gamma^4\left(\frac{1}{4}\right)}{48\pi^2} = 0.364739936
\ee
We can define $L_c(b,N)$ at a fixed $N$ and $b$ as the length where 
\be
\Omega(L_c(b,N),b,N)=0.364739936,\label{tranval}
\ee
and
\be
\lim_{N\to\infty} L_c(b,N) = L_c(b),
\ee
will be the location of the transition at infinite $N$.

Since we are working at a fixed but large $N$ in this paper, we will
define our length scale as
\be
a(b) = \frac{1}{L_c(b,N)}.\label{wilsonscale}
\ee

\subsection{Low lying fermion eigenvalues}

The eigenvalues of the massless Hermitian overlap Dirac
operator, $H_o(0)$, can 
be used to see how they scale and if they show evidence for chiral
symmetry breaking. The eigenvalues come in doubly degenerate pairs
and there is also a pairing of positive and negative eigenvalues due to
exact chiral symmetry on the lattice.
We computed all the eigenvalues of the massless
overlap Dirac operator. 

Let $0 < \lambda_k < 1$, $k=1,\cdots,N^2-1$ 
with $\lambda_k < \lambda_{k+1}$ denote all the positive
distinct eigenvalues where each eigenvalue is doubly degenerate and
each positive eigenvalue has a negative eigenvalue pair.
We can use
\be
\lambda(b) = \langle \lambda_1 \rangle\label{eigenscale}
\ee
as another choice for our length scale.

If chiral symmetry is broken, the chiral condensate sets
a scale. In particular, we expect
\be
r_k = \left\langle \frac{\lambda_1}{\lambda_k}\right\rangle\label{evratio}
\ee
to be independent of the coupling for a few low values of $k$. As $N$
increases,
we expect more $r_k$ to be independent of the coupling. 
In addition, we expect $\lambda(b)$ to approach a finite limit as
$N\to\infty$. 

\section{Single site model with massless adjoint fermions}\label{results}

Our choice of $b$ and $N$ are based on numerical feasibility.
We expect the approach to the large $N$ limit to get
slower
as we increase $b$. Since the numerical costs increase rapidly with
$N$,
we cannot make $N$ as large as we wish. Computational costs are
manageable if we choose $N$ in the range of $13$ to $25$. We will
restrict ourselves to mainly a single value of $N$, namely, $N=18$
and also provide some additional data with $N=25$ to understand finite
$N$ effects.
We have chosen the couplings in the
range
$b \in [0.32,0.70]$.
Our definition of the coupling is related to the
conventional
lattice coupling by
\be
\beta = 2bN^2.
\ee
Our range of coupling corresponds to $\beta\in[2.56,5.6]$ for SU(2)
and $\beta\in[5.76,12.6]$ for SU(3). The choice of couplings falls in
the
range of recent simulations with adjoint fermions using SU(2)
as the gauge
group~\cite{Catterall:2011zf,DeGrand:2011qd,Hietanen:2009az}
and also in simulations with fermions in the symmetric two-index representation
and SU(3) as the gauge group~\cite{Shamir:2008pb}.

In addition to these physical parameters,
we also have to choose the value of the Wilson mass parameter, $m$, in
(\ref{wilson}).
It is an irrelevant parameter but needs to be chosen in a specific
range
to realize the correct continuum limit. Based on previous studies on a
single site model with adjoint fermions~\cite{Hietanen:2010fx}, we
set $m=4$ in this paper.

\begin{table}
\center{
\begin{tabular}{lllll}
\hline
$N$ & $b$ & $\langle P\rangle $ & $a(b)$ & $N\lambda(b)$ 
\\
\hline 
18&0.32&0.6092(7)&0.4442(24)&0.0544(8)\\
18&0.35&0.6290(7)&0.4251(23)&0.0507(7)\\
18&0.40&0.6720(7)&0.3858(22)&0.0440(6)\\
18&0.45&0.7045(6)&0.3561(29)&0.0381(5)\\
18&0.50&0.7325(5)&0.3354(27)&0.0330(4)\\
18&0.53&0.7468(5)&0.3050(36)&0.0301(4)\\
18&0.55&0.7562(5)&0.2931(29)&0.0279(4)\\
18&0.57&0.7650(5)&0.2820(28)&0.0278(4)\\
18&0.60&0.7775(5)&0.2704(26)&0.0256(4)\\
18&0.65&0.7943(5)&0.2566(29)&0.0229(3)\\
18&0.70&0.8076(4)&0.2354(131)&0.0213(3)\\
25&0.40&0.6888(5)&0.3761(27)&0.0404(6)\\
25&0.45&0.7187(5)&0.3400(34)&0.0362(6)\\
25&0.50&0.7444(5)&0.3134(18)&0.0318(5)\\
25&0.55&0.7678(5)&0.2973(17)&0.0277(4)\\
25&0.60&0.7850(4)&0.2864(15)&0.0238(4)\\
25&0.65&0.8024(4)&0.2778(17)&0.0218(4)\\
\hline
\end{tabular}
\caption{\label{tab1} 
Table showing the various values of couplings where simulations were
performed with massless fermions with $N=18$ and $N=25$. The results for the
average
plaquette and the two different choices for the scales are also shown.
}}
\end{table}

\begin{figure}[ht]
\centerline{
\includegraphics[width=195mm]{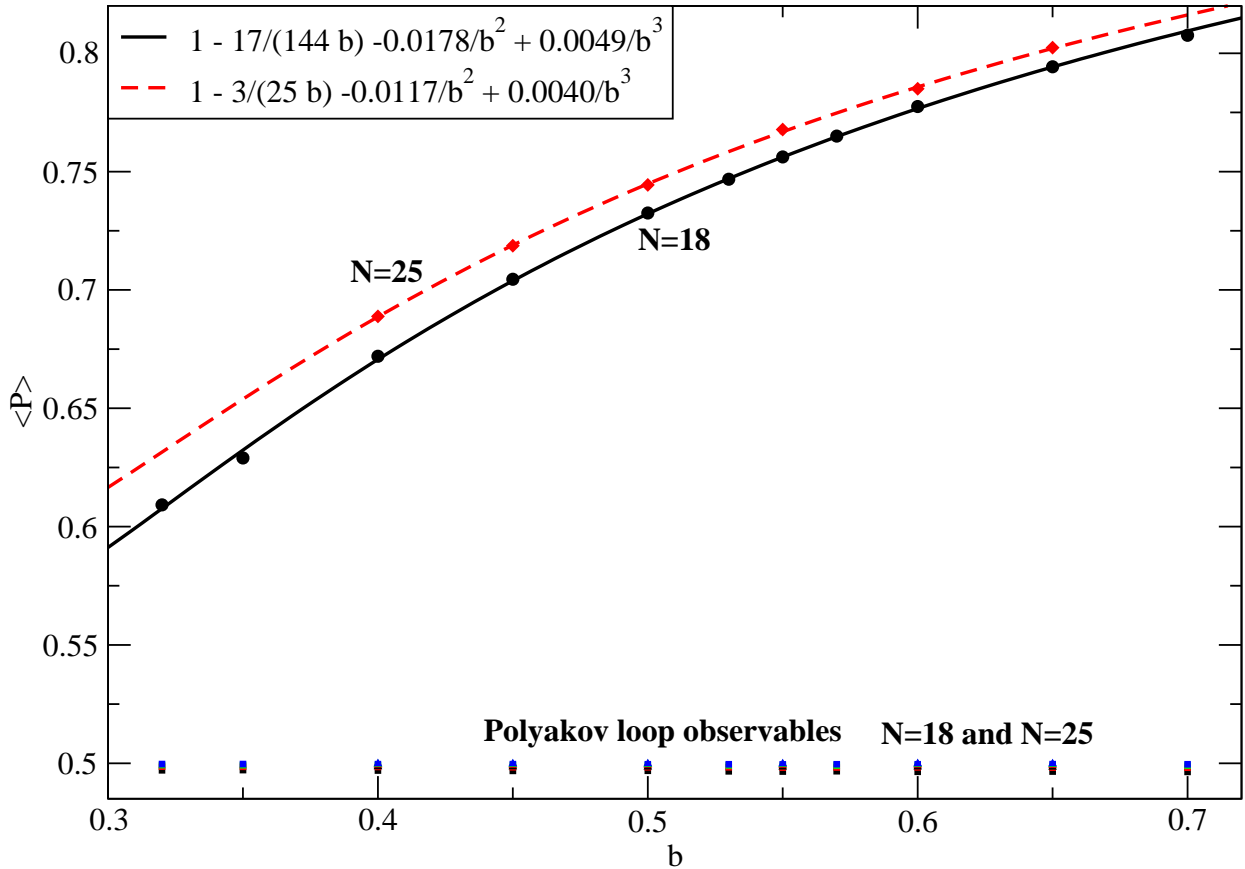}
}\caption{Average value of the plaquette along with the average values
  for
the four different Polyakov loop observables for massless fermions
at $N=18$.} \label{avgplaq}
\end{figure}

Table~\ref{tab1} 
shows the various values of couplings where
simulations were performed along with the results for the average
plaquette (c.f. (\ref{gaction})), $a(b)$
(c.f. (\ref{wilsonscale})),
and $N\lambda(b)$ (c.f. \ref{eigenscale})). 
A plot of the average
plaquette
is shown in Fig.~\ref{avgplaq}. The plaquette leads off as
$1 - \frac{N-1}{8Nb} +O(b^{-2})$ where the coefficient of $b^{-1}$ is
not affected by fermions.
 A fit of the data shows a smooth
approach to unity as $b\to\infty$. The data also shows a measure
of the fact that the eigenvalues of Polyakov loop operators
$U_\mu$ are uniformly distributed.
The four data points shown by different colored squares, correspond
to the average values of
\be
P_\mu = \frac{1}{2} \left( 1 - \frac{1}{N^2}|\Tr U_\mu|^2\right);\label{loopvar}
\ee
for $\mu=1,\cdots,4$ with $P_1 < P_2 < P_3 < P_4$ on every gauge field
configuration.
An average value of $\frac{1}{2}$ in the large $N$ limit shows that
the $Z_N$ symmetries are not broken. Our results are very close to
$\frac{1}{2}$ for both $N=18$ and $N=25$ 
and we can assume that reduction to a single site holds and
we are simulating an infinite volume theory.

\begin{figure}[ht]
\centerline{
\includegraphics[width=195mm]{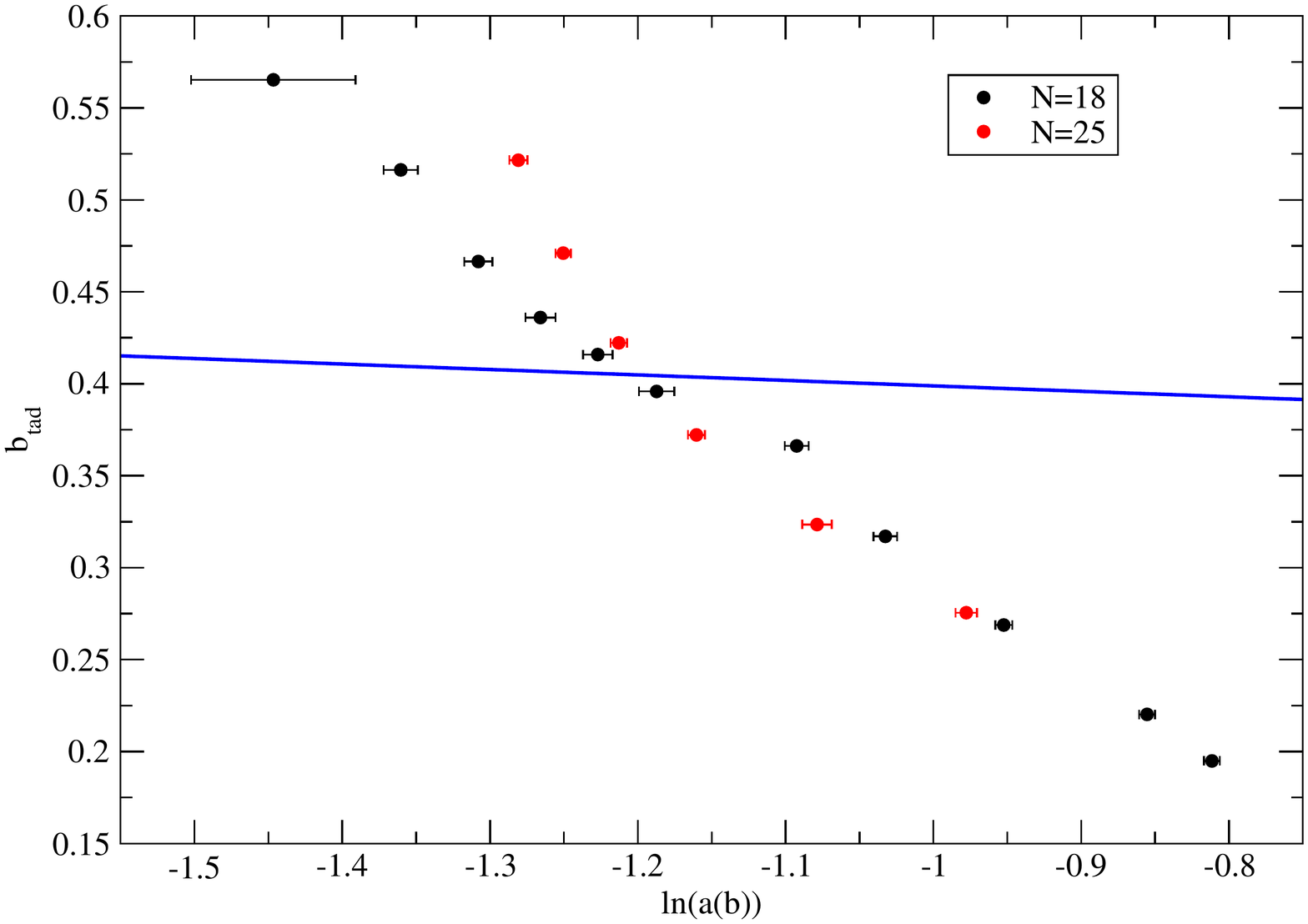}
}\caption{
Running of the tadpole improved coupling versus the logarithmic scale $\ln
a(b)$ for massless fermions at $N=18$ and $N=25$.
} \label{runningl}
\end{figure}

\begin{figure}[ht]
\centerline{
\includegraphics[width=195mm]{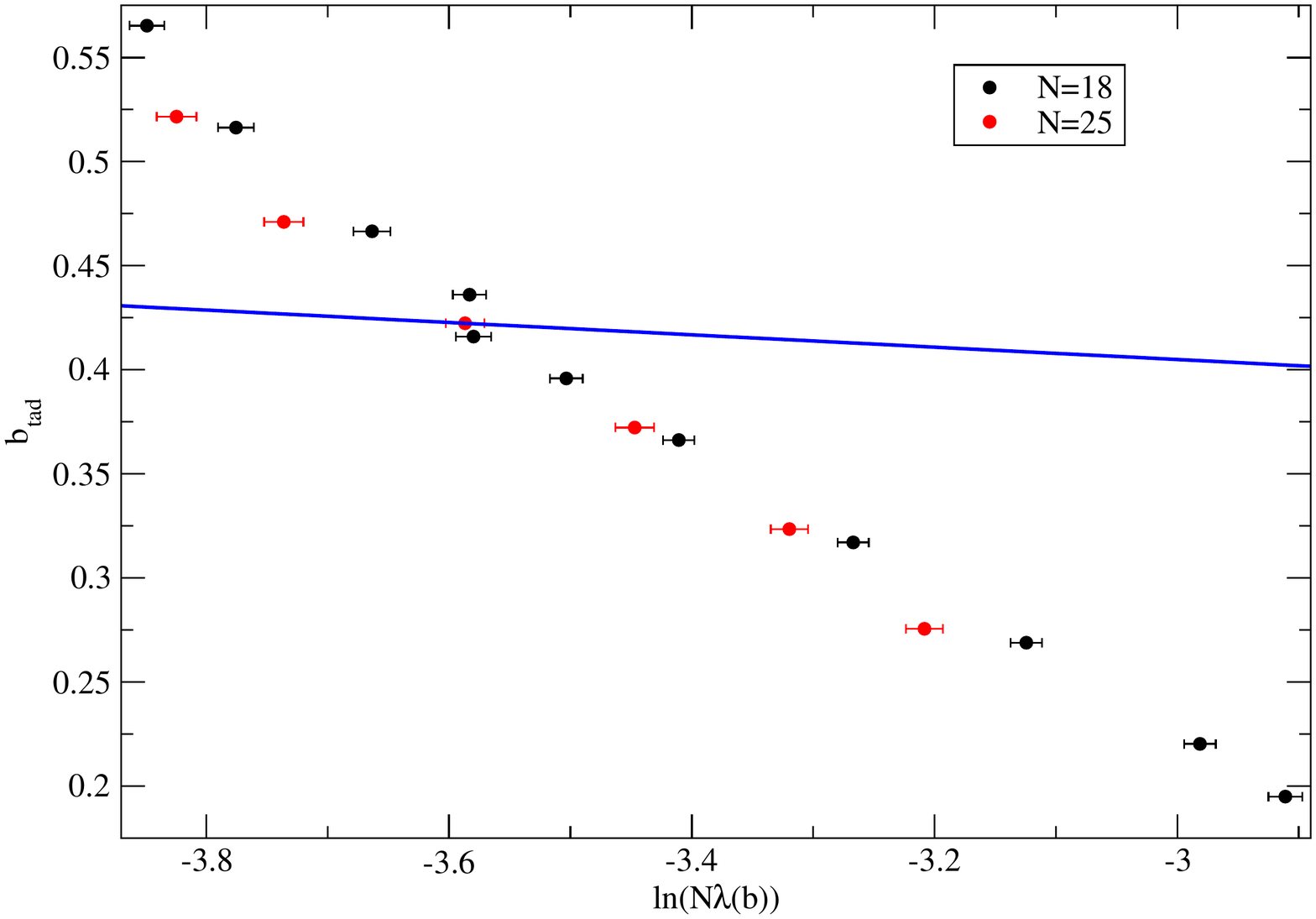}
}\caption{
Running of the tadpole improved coupling versus the logarithmic scale
$\ln \lambda(b)$ for massless fermions at $N=18$ and $N=25$.
} \label{runninge}
\end{figure}

We define 
\be
b_{\rm tad} = b \langle P\rangle,
\ee
as the tadpole improved coupling and plot the running of this coupling
versus our two logarithmic scales, $\ln a(b)$ and $\ln(N\lambda(b))$
in Fig.~\ref{runningl} and Fig.~\ref{runninge} respectively for the
data points listed in Table~\ref{tab1}. The data with errorbars are
shown with solid circles in both figures. We chose one point in the
middle
of the range as our renormalization point and the solid curve
represents
the result based on two loop perturbation theory. Clearly, there is no
agreement
and the coupling runs much faster than what is expected from two loop
perturbation theory. This indicates that we are working with lattice couplings that should
be considered as strong in spite of the fact that we used values that
would
be considered as weak in theories that do not have additional fixed points.

\begin{figure}[ht]
\centerline{
\includegraphics[width=92.5mm]{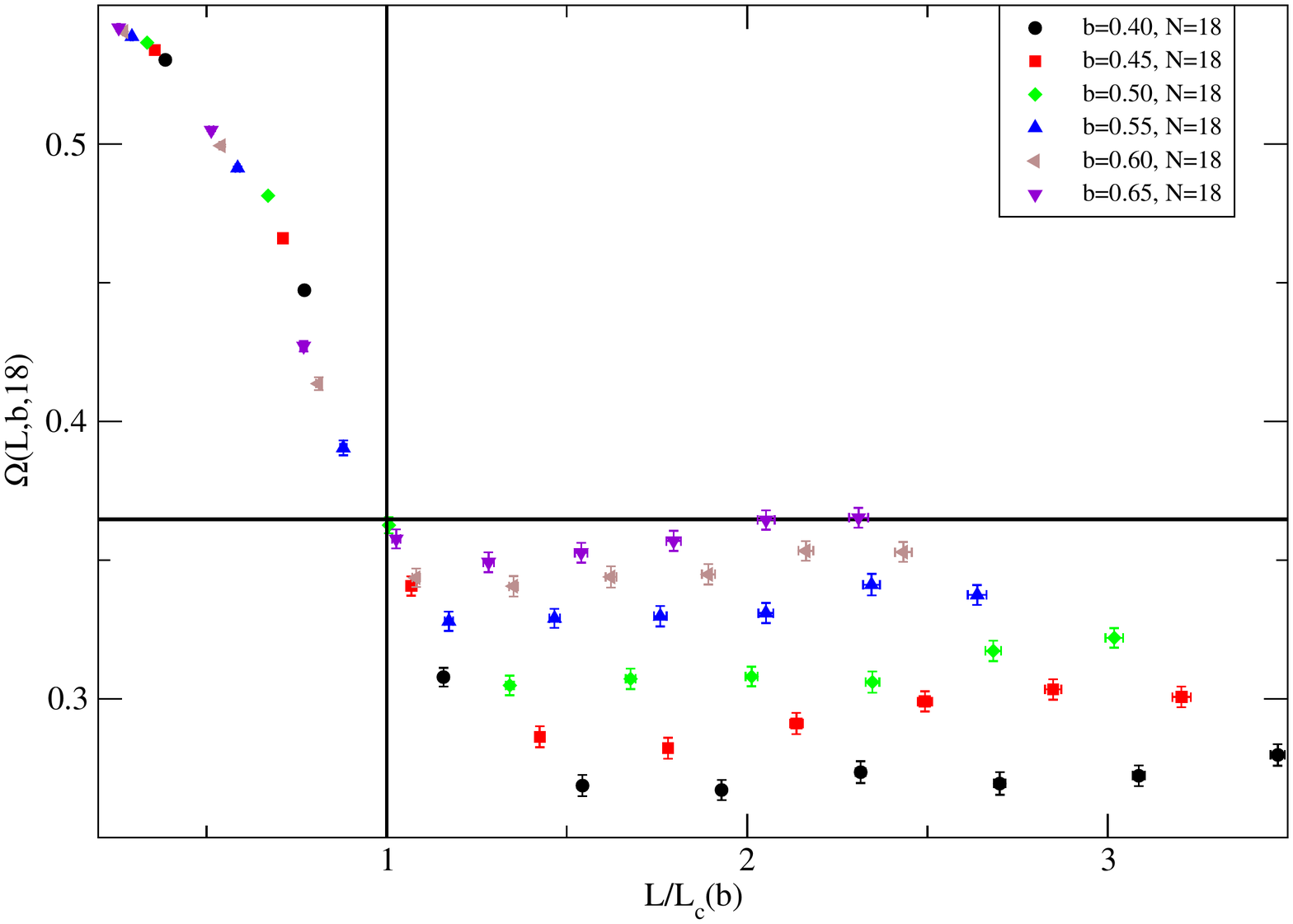}
\includegraphics[width=92.5mm]{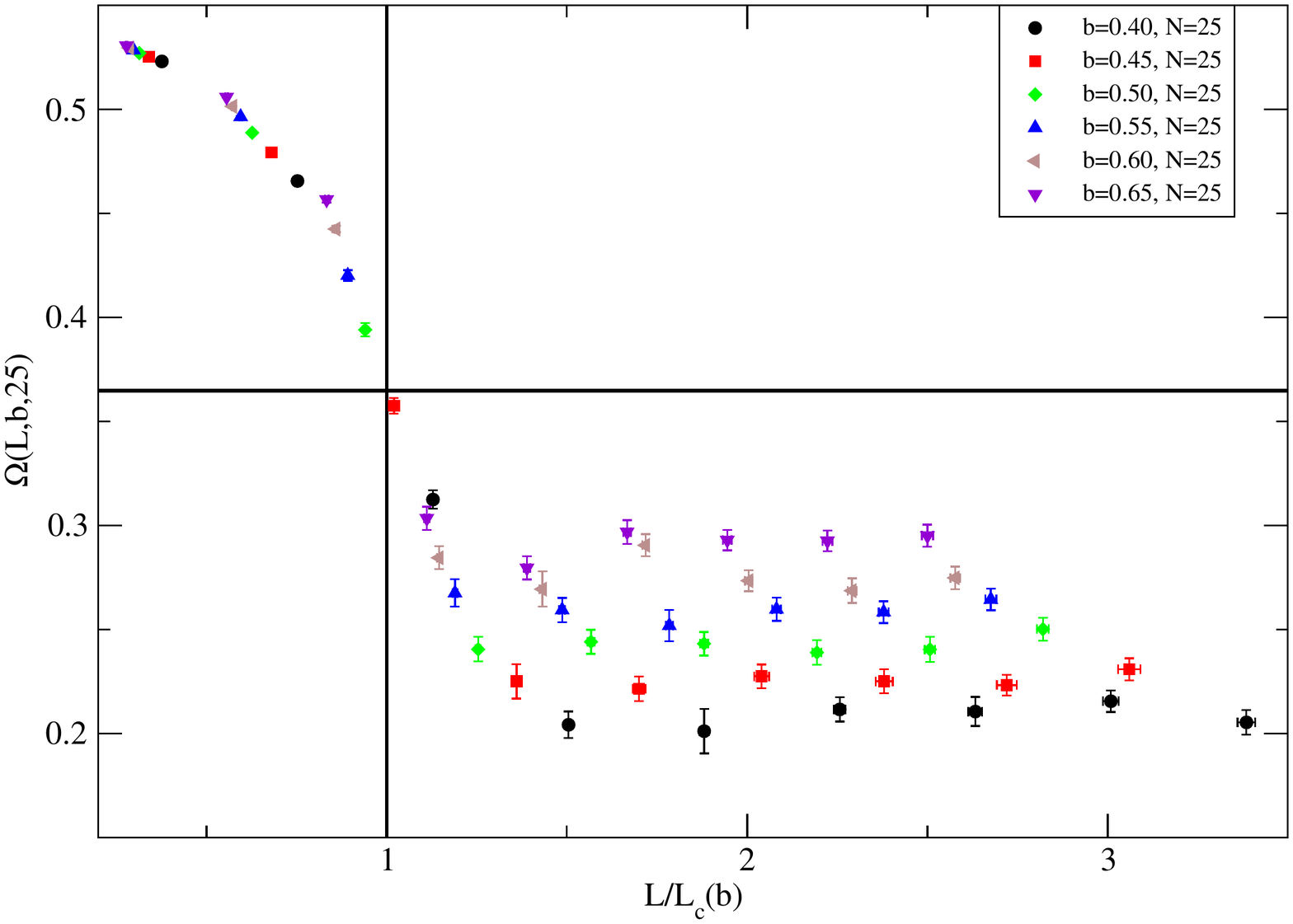}
}\caption{
Plot of the quantity, $\Omega(L,N,b)$, as a function of $L/L_c(b)$
at $N=18$  and $N=25$ for massless fermions at several different coupling.}
\label{figomega} 
\end{figure}

We end this section by presenting some details pertaining to the two scales. 
We plot $\Omega(L,b,N)$ for $N=18$ and $N=25$ in Fig.~\ref{figomega}.
We have set the x-axis to $L/L_c(b)$
  where
$L_c(b)$ is obtained using (\ref{tranval}). We see that the value for
$\Omega$ flattens out for large loops and this is a finite $N$ effect.
In addition, it flattens out at a higher value for weaker coupling and
this
is because finite $N$ effects sets in at smaller physical loop sizes
at weaker coupling. Due to these two effects, the determination of
$L_c(b)$ at weaker coupling has larger finite $N$ effects. 
We note that the finite $N$ effect get weaker at $N=25$ 
where we can perform a better estimate of
$L_c(b)$
even at $b=0.65$. The larger finite $N$ effect at weaker coupling can
also be seen in Fig.~\ref{runningl}.

\begin{figure}[ht]
\centerline{
\includegraphics[width=92.5mm]{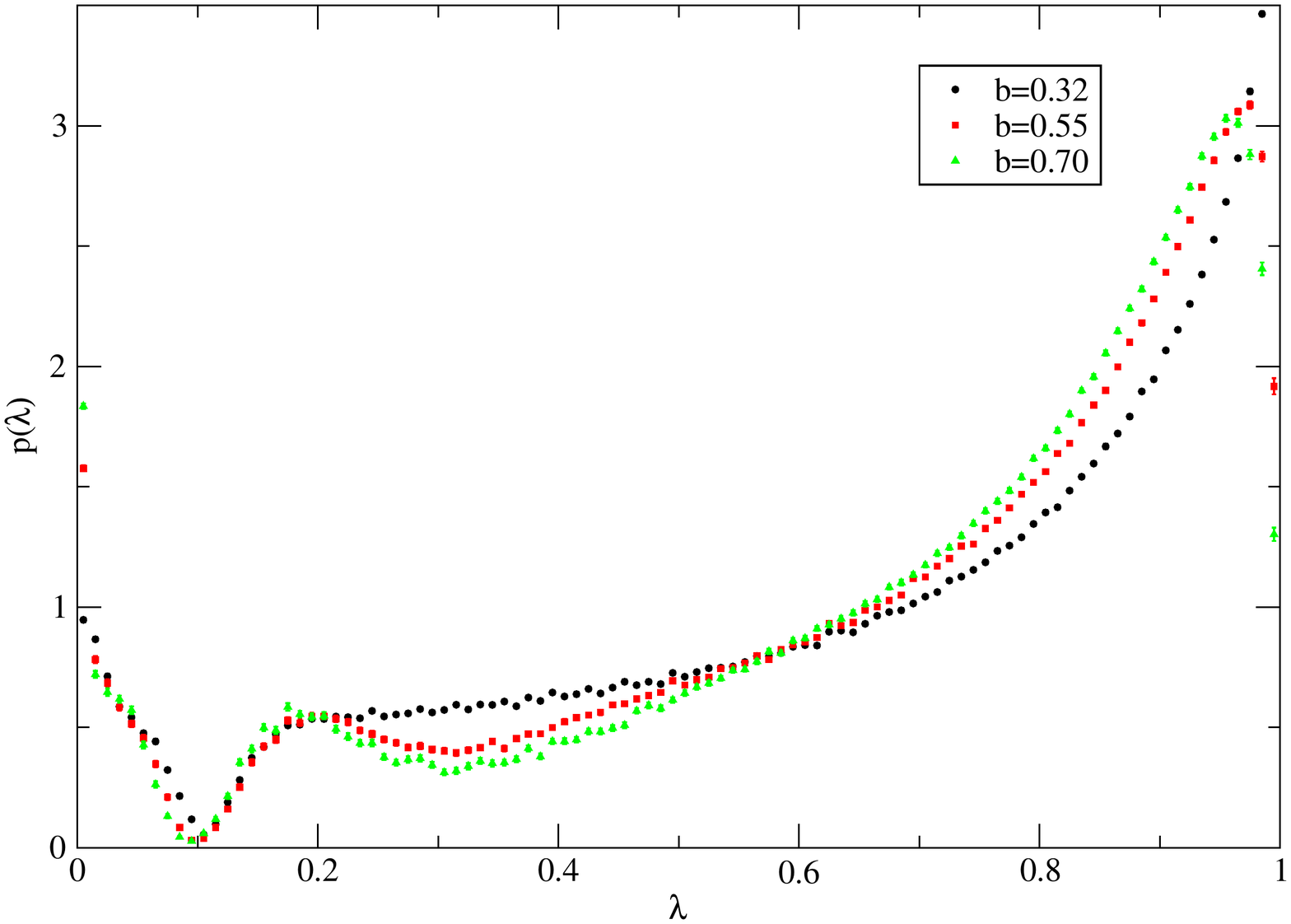}
\includegraphics[width=92.5mm]{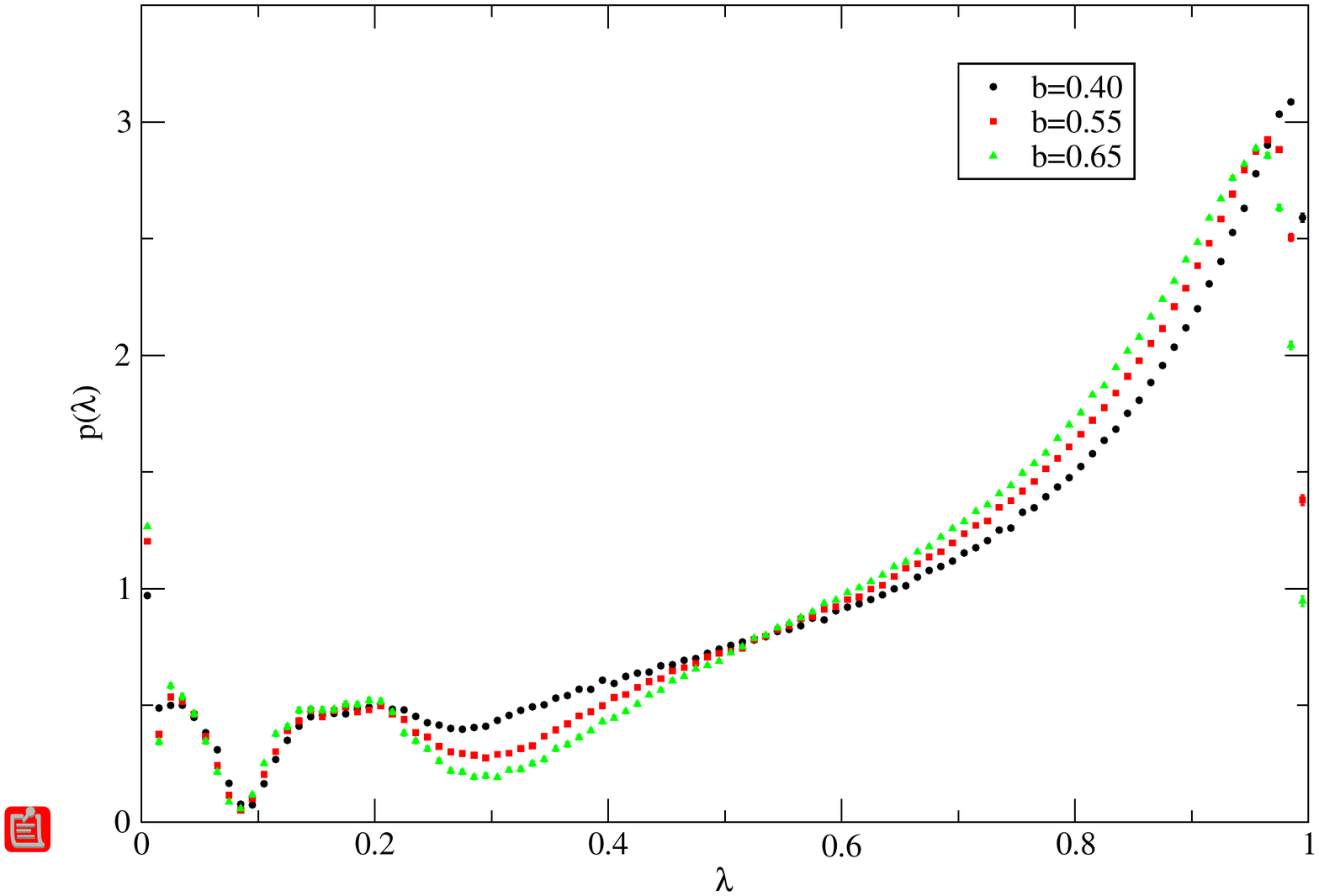}
}\caption{
The full distribution of the eigenvalues of the massless adjoint overlap Dirac operator
for three different couplings at $N=18$ and $N=25$.
}\label{odist}
\end{figure}

The complete spectrum of the distinct eigenvalues of the massless
adjoint overlap Dirac operator are shown in Fig.~\ref{odist} for three
different couplings at $N=18$ and $N=25$. All three plots show the same qualitative
behavior. We see a concentration of small eigenvalues (less than
$0.1$) followed by a bulk like distribution. We think the distribution
for $\lambda < 0.1$ is due to the would be zero modes in the 
gauge field background that is diagonal. We believe that this part of
the distribution will shows signs of chiral symmetry breaking if one
exists. If this is the case, we would expect the lowest eigenvalue to
scale like $\frac{1}{N^2}$. But we only see evidence for scaling like
$\frac{1}{N}$ in Fig.~\ref{runninge} where a plot with
$\ln(N\lambda(b))$
on the x-axis show good agreement between $N=18$ and $N=25$. Therefore, we do not yet see evidence for chiral
symmetry breaking.

\begin{figure}[ht]
\centerline{
\includegraphics[width=92.5mm]{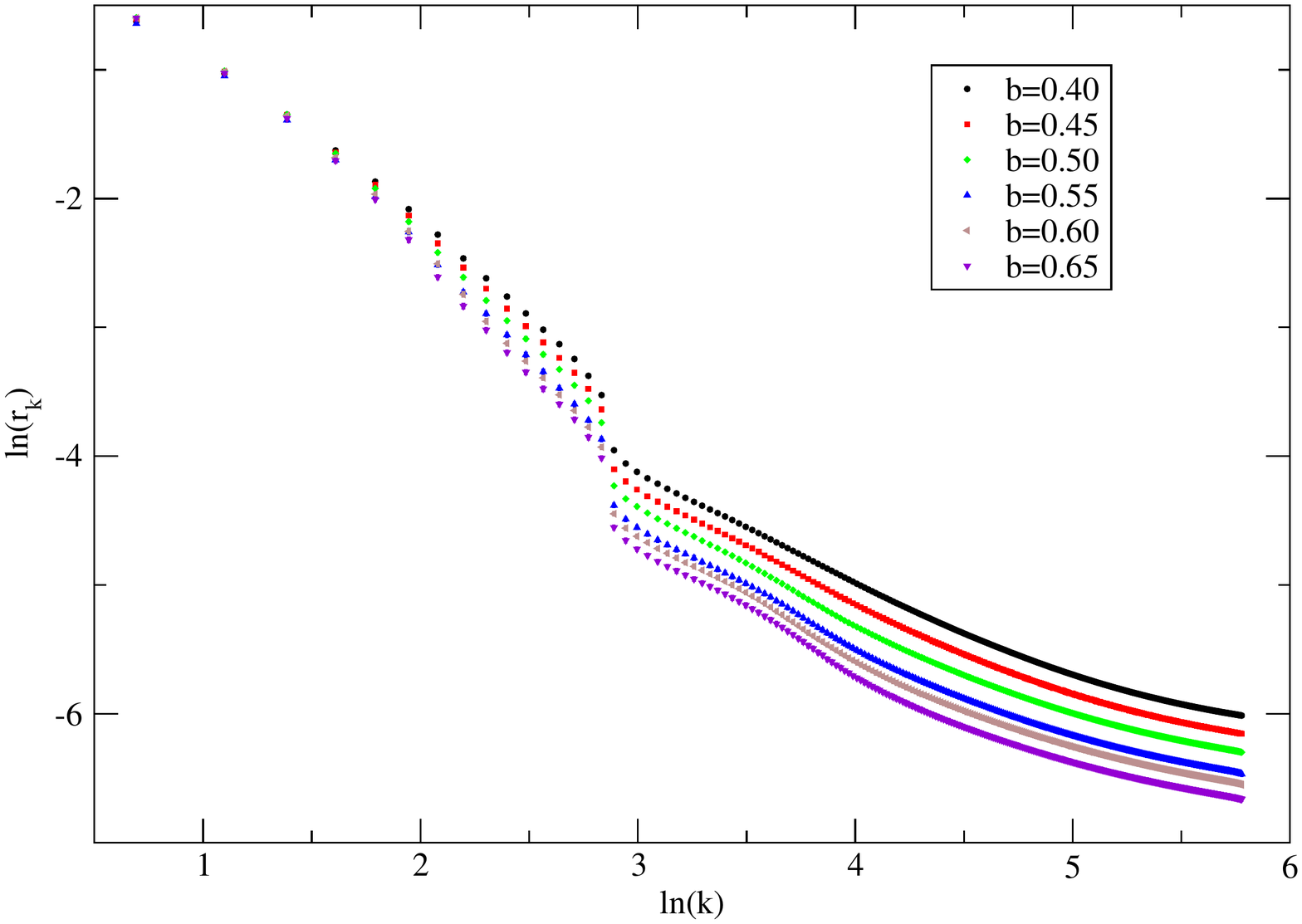}
\includegraphics[width=92.5mm]{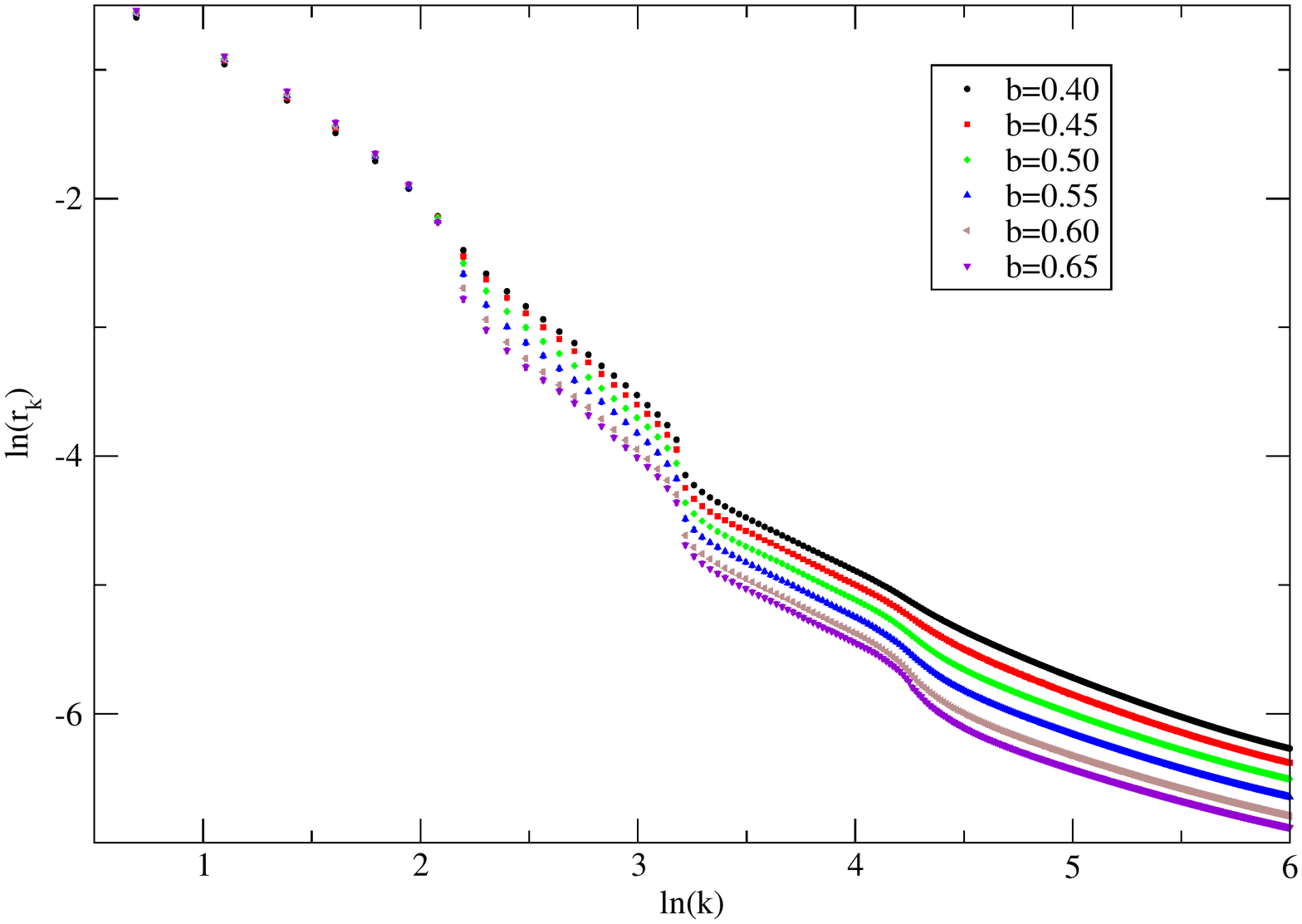}
}\caption{
The ratios of the eigenvalues of the massless adjoint overlap Dirac operator
are shown in log-log plot for all couplings common to $N=18$ and
$N=25$ in Table.~\ref{tab1}.
}\label{ratioplot}
\end{figure}

The plot of $r_k$ as defined in (\ref{evratio}) versus $k$ is shown in
a log-log plot in Fig.~\ref{ratioplot}. Here again, one sees a
separation between the low eigenvalues (the would-be zero modes in
a diagonal background) and the bulk. Furthermore, the ratios do not
change much with coupling for $k<5$ for $N=18$ and $k<7$ for $N=25$
showing that the finite $N$ effect is of order $\frac{1}{N}$.

\section{Speculations}\label{nonanalytic}

How does one see the effect of an infra-red fixed point in a
lattice computation? According to the two loop beta function, the zero
occurs at
\be
b_*(f) = -\frac{b_1}{b_0} = \frac{1}{8\pi^2}\frac{16f - 17}{11-4f}.
\ee
If $f$ is close to the upper limit of $\frac{11}{4}$, the zero occurs
at a value of the coupling that can be considered as
perturbative. Different choices for $a(b)$ will not matter in a study
of the infra-red fixed point.
One should see numerical 
evidence for a very rapid change of $a(b)$ versus $b$ close to
$b_*(f)$ indicative of a 
zero of the beta function.  
The critical values for $f=\frac{3}{2},2$ and $\frac{5}{2}$ are
$b_*\left(\frac{3}{2}\right)=\frac{1.4}{8\pi^2}$,
$b_*(2)=\frac{5}{8\pi^2}$
respectively. $b_*\left(\frac{5}{2}\right)=\frac{23}{8\pi^2}$.
None of these would be considered
perturbative and it is quite likely that the study of the
infra-red fixed point is strongly affected by the choice of $a(b)$.
In particular, it is quite possible that the location of the infra-red
fixed point depends on the choice of $a(b)$ and there is even a possibility
that the existence of an infra-red fixed point on either side of the
lower boundary of $f=\frac{17}{16}$ depends on the choice of $a(b)$.
We have set $f=1$ in this study and we clearly do not see perturbative
behavior
as discussed in Sec.~\ref{results}.

In order to understand the behavior of the
running coupling better, we consider beta functions of the form
\bea
{\bf A:} & \beta(b) & = - b_0\left( 1- \frac{b_I}{b}\right) \cr
{\bf B:} & \beta(b) & = - b_0\left( 1- \frac{b_I}{b} \right)\left( 1-
  \frac{b_U}{b} \right) \cr
{\bf C:} & \beta(b) & = - b_0\left[ \left( 1- \frac{b_*}{b}
  \right)^2+\frac{\alpha^2}{b^2} \right]\cr
{\bf D:} & \beta(b) & = \cases{ 
- b_0\left|1- \frac{b_*}{b}
  \right|^{p_+} & for $b>b_*$ \cr
- b_0 s \left|1- \frac{b_*}{b}
  \right|^{p_-} & for $b< b_*$ \cr}\label{betacases}
\eea
where it is assumed that all parameters except $s$ are positive and we also
assume that $b_I > b_U$. The parameter $s$ could be positive or negative.

\begin{figure}[ht]
\centerline{
\includegraphics[width=195mm]{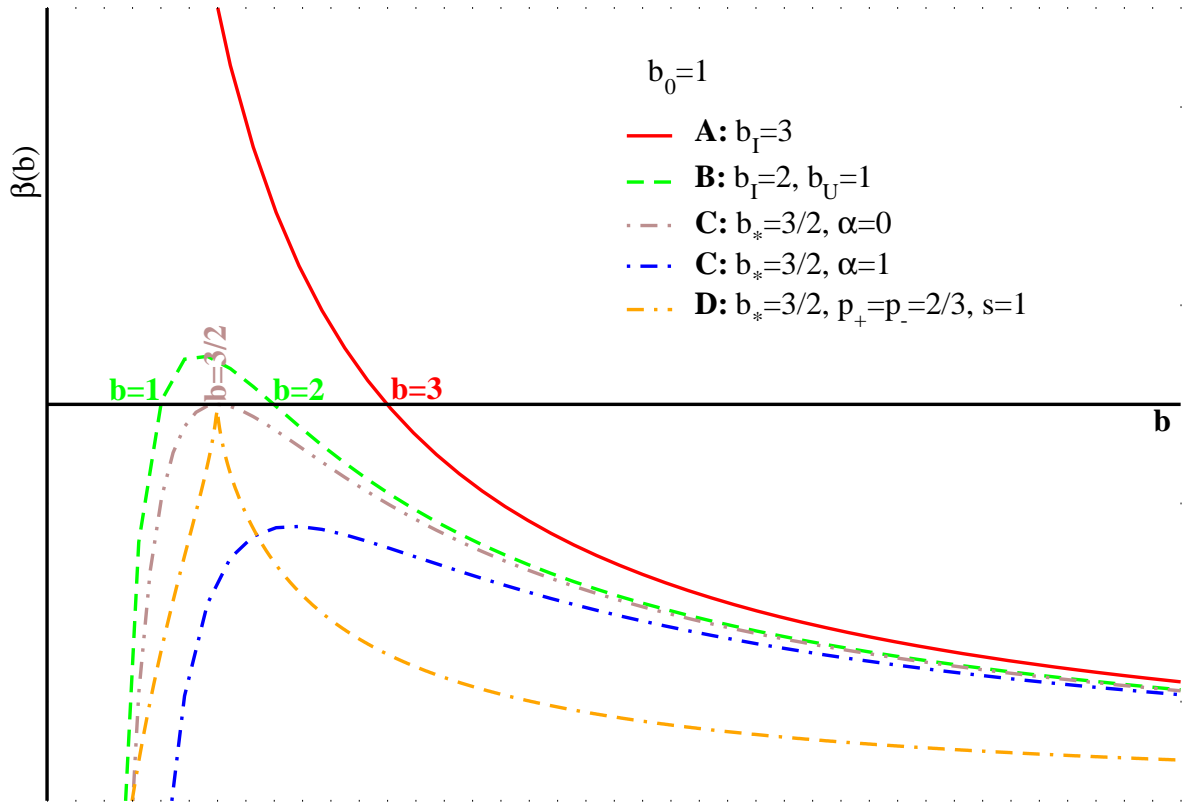}
}\caption{Different cases for the beta function are shown.} \label{fig1}
\end{figure}

\begin{figure}[ht]
\centerline{
\includegraphics[width=195mm]{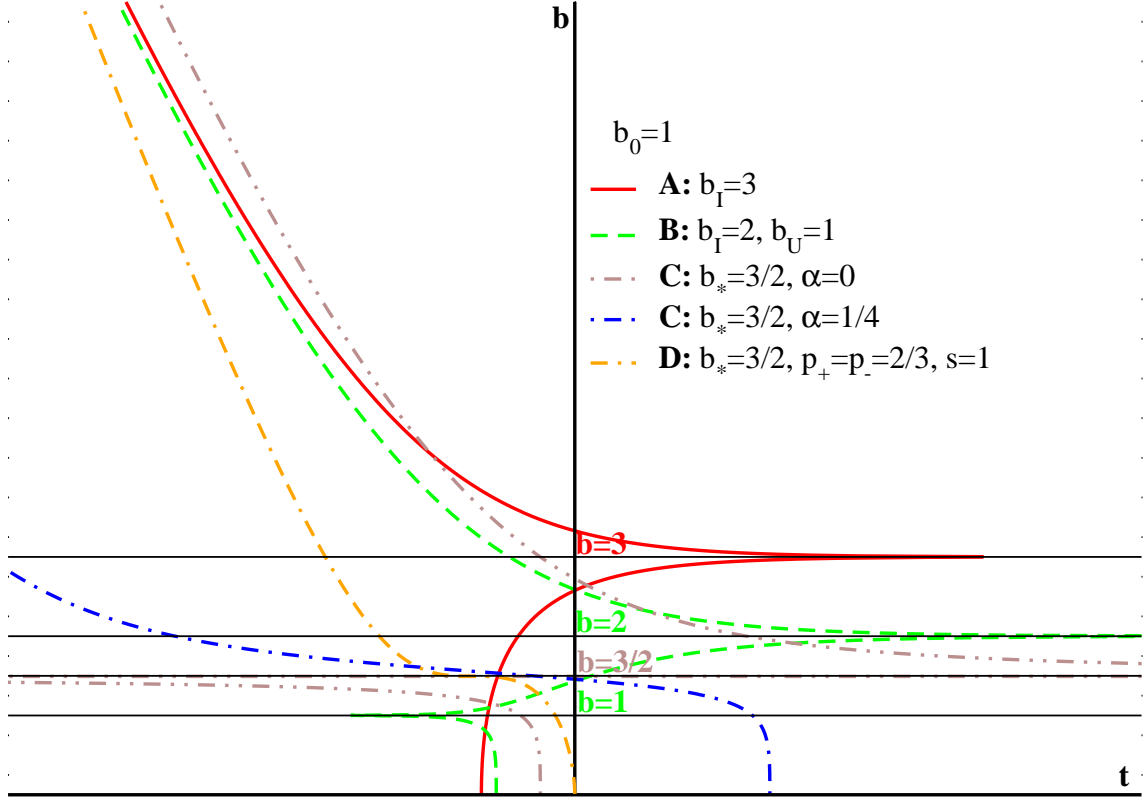}
}\caption{The running of the coupling with the scale is shown for the
  different cases of the beta function.} \label{fig2}
\end{figure}

\begin{figure}[ht]
\centerline{
\includegraphics[width=195mm]{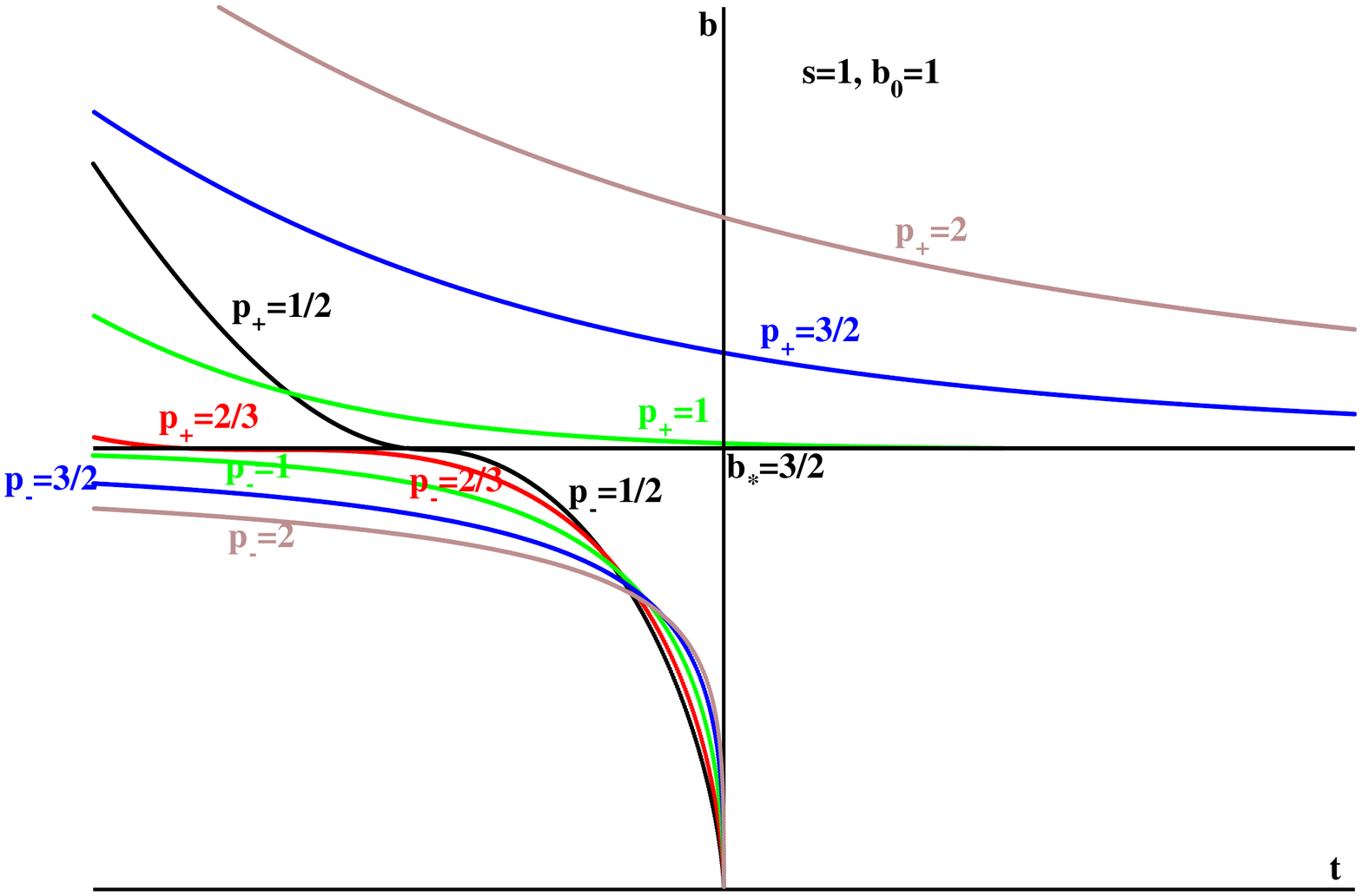}
}\caption{The running of the coupling with the scale is shown for case
  {\bf D} with 
  different choices for $p_\pm$ with $s=1$ and $b_0=1$.} \label{fig3}
\end{figure}

\begin{itemize}
\item
Case ${\bf A}$ is the two loop beta function written for the
case where it has a zero. $b_I$ is the location of the infra-red fixed
point.
The beta function is shown in solid red in Fig.~\ref{fig1} and the
a plot of the coupling with the scale is shown in solid red in
Fig.~\ref{fig2}. The logarithmic scale goes to positive infinity as
$b\to b_I$. In order to define the continuum limit, we need to take
the limit $t\to-\infty$ and this is achieved by taking $b\to\infty$
in the usual manner. This is what one expects to see if nothing
occurs beyond what is expected in two-loop perturbation theory.
\item
Cases ${\bf B}$ and ${\bf C}$ have an
additional
$\frac{1}{b^2}$ term and were discussed in~\cite{Kaplan:2009kr}.
\begin{itemize}
\item 
The beta function and the dependence of coupling on the scale for case
${\bf B}$   
are shown in dashed green in Fig.~\ref{fig1} and Fig.~\ref{fig2}  
respectively.  
The logarithmic scale goes to positive infinity at $b=b_I$ and goes to
negative
infinity at $b=b_U$ which is an ultra-violet fixed point. In addition
to defining a continuum limit as $b\to\infty$, we can also define
a continuum theory as $b\to b_U$. It is likely that the location of
$b_U$
depends on the choice of $a(b)$. If we assume that a $b_U$ exists for
all choices of $a(b)$, then we can define a continuum theory by taking
a limit $b\to b_U$ that is operator dependent. 
\item
The case
${\bf B}$ with $b_I=b_U=b_*$ is same as case ${\bf C}$ with $\alpha=0$.
 The beta function and the dependence of coupling on the scale for this case
are shown in dot-dot-dashed brown in Fig.~\ref{fig1} and Fig.~\ref{fig2}
respectively.
The limit of $b\to b_*$ from above results in the logarithmic scale
approaching positive
infinity and 
the limit of $b\to b_*$ from below results in the logarithmic scale
approaching negative infinity. A continuum theory can be defined in the
limit of $b\to b_*$ from below. As before, $b_*$ is expected to depend
on the
choice of $a(b)$.
\item
The beta function and the dependence of coupling on the scale for case
${\bf C}$ with $\alpha>0$  
are shown in dot-dashed blue in Fig.~\ref{fig1} and Fig.~\ref{fig2}  
respectively.  
In this case, there is a region around $b=b_*$ where the coupling {\sl
  walks}.
How {\sl far it walks}, depends on $\alpha$ and the range $[t_I,t_U]$,
where the coupling essentially remains a constant, grows as $\alpha$
decreases. As before, $b_*$ is expected to depend on the choice of
$a(b)$. The continuum limit is defined as $b\to\infty$.
\end{itemize}
\item In all cases discussed above, the location of $b_I$ and/or $b_U$
  depends
on the choice of $a(b)$. 
Case {\bf D} is different in this respect. If the beta function has a
zero that is non-analytic, we expect the location of the zero to not
depend on $a(b)$. In other words, if one observable shows non-analytic
behavior at some coupling, all observables are expected to show
non-analyticity at the same coupling.
If $s=-1$ and $p_+=p_-=1$, we recover case
  {\bf A}. If $s=1$ and $p_+=p_-=2$, we recover case {\bf C} with
    $\alpha=0$.
The beta function and the dependence of coupling on the scale for
$s=1$ and $p_+=p_-=\frac{2}{3}$ 
are shown in dot-dash-dashed orange in Fig.~\ref{fig1} and Fig.~\ref{fig2} 
respectively. 
Like in case {\bf C}, for this choice of parameters, there is a region around $b=b_*$ where the
coupling {\sl walks}. Whether there is a infra-red/ultra-violet fixed
point at $b=b_*$ depends on the choice of $p_\pm$ as can be seen from Fig.~\ref{fig3}.
\begin{itemize}
\item If $p_-< 1$, the scale change from $b=0$ to  $b=b_*$ is
  finite. In such a situation, we cannot define a
  continuum theory by taking $b\to b_*$ from below.
\item If $p_- >1$, the  logarithmic scale goes to negative infinity as
  $b\to b_*$ from below and $b_*$ is an ultra-violet fixed point from
  below.
We can define a continuum theory in this
  limit.
\item if $p_+ < 1$, the logarithmic scale starts out at a finite value
  for $b=b_*$
  and goes down to negative infinity as $b\to\infty$. 
A continuum
  theory can be defined as $b\to\infty$.
\item If $p_+ > 1$, $b=b_*$ is an infra-red fixed point from above and
the logarithmic scale goes to positive infinity as $b\to b_*$ from
above.
A continuum
  theory can be defined as $b\to\infty$.
\end{itemize}
\end{itemize}

As the maximum of $\beta(b)$ moves away from zero in case {\bf C}, the
range where the {\sl coupling runs slowly} shrinks. As the power,
$p_\pm$, goes below unity in case {\bf D}. the range where the
{\sl coupling runs slowly} shrinks. In such cases, it will be
difficult to see evidence for a slow running of the coupling. Instead,
one
will see a fast running of the coupling on either side of the maximum
in case {\bf C} or the location of the singular point in case {\bf D}.
The numerical data shown in Fig.~\ref{runningl} and
Fig.~\ref{runninge} is close to what is seen in case {\bf C} or case
{\bf D} for couplings around $b_*$.  Since we see do not see agreement
with perturbation theory and we see a running that is significantly
faster, the couplings we are using on the lattice are probably on
either side
of $b_*$ and $f=1$ probably corresponds to a case where the region of
{\sl slow running of coupling} is very small. Clearly, our data is not precise
enough to distinguish between either of these cases. But our data
suggests
the presence of a $b_*$.

\section{Conclusions}

The single site model of a  large $N$ gauge theory coupled to massless
adjoint fermions was numerically studied in this paper. We have
studied
this model with a single flavor of adjoint fermion
numerically using the Hybrid Monte Carlo algorithm and
pseudofermions. We studied the running coupling using two different
choice
of scales and did not find agreement with two-loop perturbation theory
at intermediate values of the tadpole improved coupling. The
two different choices of scales were
the transition from weak to strong coupling and the lowest eigenvalue
of
the massless overlap Dirac operator. This is the main result of our
paper.

Since one flavor of adjoint fermion is close to $\frac{17}{16}$ which
is the lower bound for the number of flavors for a perturbative zero
of the beta function, we speculate on the possibility that a near zero of
the beta function might be the cause for our result being not in
agreement
with perturbation theory. 
Our numerical data cannot
show with definiteness that there is a near-zero of the beta function
but the behavior suggests such a possibility.

The results presented in this paper are exploratory in
nature and future simulations with different values of $f$ will give a
clearer physics picture. However,
the work lays the foundation for the careful study of ultra-violet/infra-red
fixed  points in matrix models that mimic large $N$ gauge theories
coupled to adjoint fermions. We have the ability to treat the number
of fermion flavors
as a real number in the matrix model and study the presence of
singular behavior in the associated beta function. It is likely that
the
behavior at the singular point, results in it being an
ultra-violet/infra-red fixed point for some range of fermion flavors.
The numerical procedure developed in this paper for the case
of a single Dirac flavor paves the way for future numerical studies
of the matrix model with varying number of flavors.

\begin{acknowledgments} 
R.N. acknowledges partial support by the NSF under grant number
PHY-0854744.  R.N. would like to thank Erich Poppitz for several
useful discussions.
The numerical calculations presented in this work have been performed
on the Horseshoe6 cluster at the University of Southern Denmark (SDU) funded by the 
Danish Center for Scientific Computing for the project "Origin of Mass" 2009/2010.
\end{acknowledgments}

\end{document}